\documentclass[final,5p,times,twocolumn]{elsarticle}
\usepackage{color}
\usepackage{graphicx}% Include figure files
\usepackage{bm}% bold math
\usepackage{booktabs}
\usepackage{lineno}
\def\Journal#1#2#3#4{{#1}\, {\bf #2}, #3 (#4)}
  
\def\NIMA{Nucl.\ Instrum.\ Meth.\ A}
\def\NPA{Nucl.\ Phys.\ A}
\def\NPB{Nucl.\,Phys.\,B}
\def\PLB{Phys.\ Lett.\ B}
\def\PRL{Phys.\ Rev.\ Lett.}
\def\PRD{Phys.\,Rev.\,D}

\def\JHEP{JHEP}

\begin{document}

\begin{frontmatter}

%%%\preprint{}

\title{Cross Sections and Transverse Single-Spin Asymmetries in Forward Jet Production from Proton 
Collisions at $\sqrt{s}=500\,$GeV}

\address[label1]{Brookhaven National Laboratory, Upton, New York 11973, USA}
\address[label2]{Christopher Newport University, Newport News, Virginia 23606, USA}
\address[label3]{University of California, Berkeley, California 94720, USA}
\address[label4]{Institute of High Energy Physics, Protvino 142281, Russia}
\address[label5]{Thomas Jefferson National Accelerator Facility, Newport News, Virginia 23606, USA}
\address[label6]{Shandong University, Jinan, Shandong 250100, China}
\address[label7]{University of Virginia, Charlottesville, Virginia 22903, USA}
\address[label8]{College of William and Mary, Williamsburg, Virginia 23187, USA}
\address[label9]{University of Zagreb, Zagreb, HR-10002, Croatia}
\address[label10]{Norfolk State University, Norfolk, Virginia 23504, USA}
\address[label11]{Department of Theoretical Physics, University of the Basque Country UPV/EHU, 48080 Bilbao, Spain}
\address[label12]{IKERBASQUE, Basque Foundation for Science, 48011 Bilbao, Spain}

\author[label1]{L.~C.~Bland}
\author[label2]{E.~J.~Brash}
\author[label3]{H.~J.~Crawford}
\author[label4]{A.A.~Derevschikov}
\author[label1]{K.~A.~Drees}
\author[label3]{J.~Engelage}
\author[label1]{C.~Folz}
\author[label5]{M.~K.~Jones}
\author[label3]{E.~G.~Judd}
\author[label6,label1]{X.~Li}
\author[label7]{N.~K.~Liyanage}
\author[label1]{Y.~Makdisi}
\author[label4]{N.~G.~Minaev}
\author[label2]{R.~N.~Munroe}
\author[label4]{L.~Nogach}
\author[label1]{A.~Ogawa}
\author[label8]{C.~F.~Perdrisat}
\author[label3]{C.~Perkins}
\author[label9]{M.~Planinic}
\author[label10]{V.~Punjabi}
\author[label11,label12]{G.~Schnell}
\author[label9,label1]{G.~Simatovic}
\author[label1]{T.~G.~Throwe}
\author[label11]{C.~Van Hulse}
\author[label4]{A.~N.~Vasiliev}

\address{\rm\normalsize(A$_N$DY Collaboration)$^*$}
%\address{(A$_N$DY Collaboration)\corref{cor1}}
\cortext[cor1]{\it{URL: www.andy.bnl.gov}}

%%%\collaboration{A$_N$DY Collaboration}\homepage{www.andy.bnl.gov}\noaffiliation

\date{\today}

\begin{abstract}
Measurements of the production of forward jets from transversely polarized proton collisions 
at $\sqrt{s}=500\,$GeV conducted at the Relativistic Heavy Ion Collider (RHIC) 
are reported. Our measured jet cross section is consistent with hard scattering 
expectations. Our measured analyzing power for forward jet production is small 
and positive, and provides constraints on the Sivers functions that are related 
to partonic orbital angular momentum through theoretical models. 

\vspace{0.2cm}
\end{abstract}

%%%\pacs{12.38.Qk,13.87-a,13.88+e}
%%\keywords{low-$x$, saturation, Color Glass Condensate, particle
%%                production, shadowing} 

%\begin{keyword}
%\end{keyword}

%%%\maketitle

\end{frontmatter}

%%\begin{linenumbers}

The proton is a building block of matter, which is itself built from
elementary quarks and gluons.  Our understanding of the structure of the 
proton has become increasingly sophisticated since the advent of Quantum 
Chromodynamics (QCD), and a reason for this has been the quest to understand 
how the proton gets its intrinsic spin from its constituents.  The present 
view is that quark or gluon orbital angular momentum (OAM) makes important 
contributions to the proton spin~\cite{kfliu}.
Early indications of this came from large analyzing powers
($A_N$), also known as transverse single-spin asymmetries (SSA),  
measured in the production of charged and neutral pions in
collisions of transversely polarized protons at center-of-mass 
energy $\sqrt{s}=20\,$GeV~\cite{E704}.  The observable $A_N$ is the amplitude of
the spin-correlated azimuthal modulation of the produced particles.
A large $A_N$ is not expected for pions produced with sufficient
transverse momentum ($p_T$) in collinear perturbative QCD (pQCD) at
leading twist, due to the chiral properties of the theory~\cite{Ka78}.
Measurements of large $A_N$ for pion production at large Feynman-$x$ 
($x_F=2p_z/\sqrt{s}$, where $p_z$ is the pion longitudinal momentum in the center-of-mass frame) 
prompted theorists to
introduce spin-correlated transverse momentum ($k_T$) in either the
initial state (Sivers effect~\cite{Si90}) or the final state (Collins
effect~\cite{Co93}). For inclusive pion production, these effects
cannot be disentangled.  In contrast, in measurements of jets, defined 
as a collimated multiplicity of energetic baryons and mesons that are produced 
in high-energy collisions, contributions to $A_N$ from final-state fragmentation 
are absent and hence information about the scattered quark or gluon can be 
inferred directly.  In particular, $A_N$ for jet production, direct photons, 
or Drell-Yan processes is expected to arise only from the Sivers effect.
The initial-state spin-correlated $k_T$ is related by models~\cite{Bu06} 
to quark and gluon OAM.

Cross sections for pion production at large $x_F$ in $p^{\uparrow}+p$
collisions at $\sqrt{s}\le20\,$GeV~\cite{E704,lowE} are much larger
than naive pQCD expectations.  This resulted in skepticism that pion
production in these kinematics is from hard-scattering processes.  Theoretical interest 
in understanding $A_N$ for pion production has been revived by recent 
measurements~\cite{STARpi0,BRAHMS} at $\sqrt{s}\ge62\,$~GeV, where 
cross sections~\cite{STARcs} are in agreement with pQCD.  
Furthermore, measurements of $A_N$ for pion production in 
$p^{\uparrow}+p$ collisions at $\sqrt{s}\ge62\,$~GeV have been concurrent 
with measurements of transverse SSA in semi-inclusive deep-inelastic
scattering (SIDIS)~\cite{SIDIS} where an electron or muon is inelastically
scattered from a proton, whose spin is transverse to the lepton
beam.  Meson fragments of the struck quark are found to have
spin-correlated azimuthal modulations, whose amplitudes are understood
by the Sivers and Collins effects, introduced to explain $A_N$ for
$p^{\uparrow}+p\rightarrow{\pi}+X$.  
An alternative and complementary theoretical approach based on 
collinear  factorization~\cite{ETQS} predicts $A_{N}$ involving twist-3 
multi-parton correlations~\cite{twist3SSA}, and is expected to be related 
to the Sivers and Collins functions via transverse-momentum 
moments. However, an attempt at linking the different approaches 
using data from SIDIS and $p^{\uparrow}+p\rightarrow{\pi}+X$
yielded a mismatch in the sign~\cite{sign-mismatch}. 
Most recently, theory has proposed that transverse SSA in $p^\uparrow+p$ 
receive large contributions from fragmentation~\cite{KKMP}.
Thus a consistent understanding of all transverse SSA in hard 
scattering processes is not yet within our grasp, but would greatly benefit 
from measurements of $A_N$ for jet production in $p^{\uparrow}+p$
collisions, since it receives no contributions from spin-dependent
fragmentation effects.

\begin{figure}
  \centering
  \includegraphics[height=2.48in]{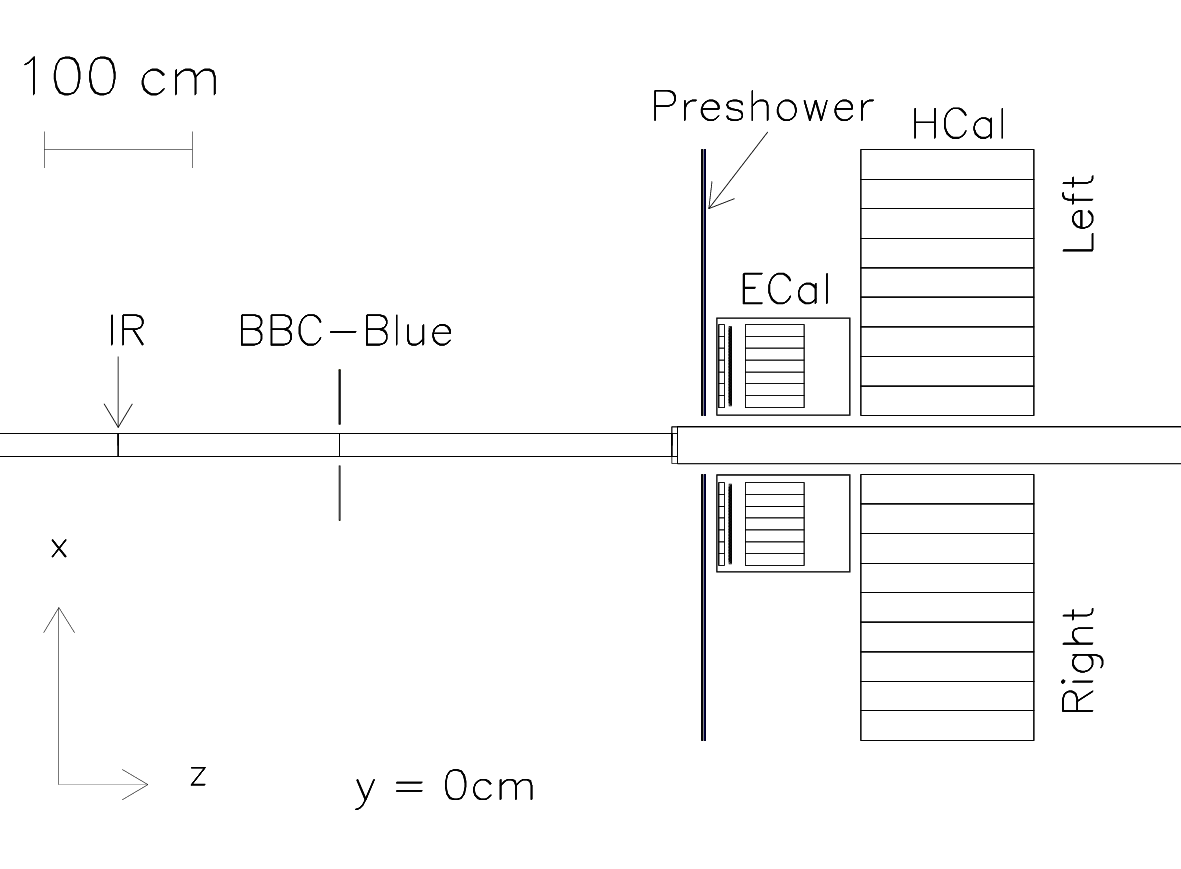}
  \caption{A top view from the GEANT model of A$_N$DY configuration 
  for the 2011 run. The Blue beam travels in the positive $z$ direction, and 
  Yellow beam in the opposite direction. IR indicates the center of 
  the collision region.}
  \label{setup}
\end{figure}

In this Letter, we report first measurements of cross sections and
$A_N$ for forward jet production in $p^{\uparrow}+p$
collisions at $\sqrt{s}=500\,$GeV.  The measurement was conducted with
the A$_N$DY detector at the 2 o'clock interaction region (IP2) of
RHIC at Brookhaven National Laboratory.  The primary detector components 
were two mirror-symmetric hadron calorimeter (HCal) modules that were 
mounted to face the ``Blue'' beam (the ``Yellow'' beam 
travels in the opposite direction) for the 2011 
and 2012 RHIC runs. The HCal spanned the pseudorapidity interval 
$2.4<\eta<4.0$, a region that is well shielded from single beam backgrounds 
by the cryostats of the ring magnets.  A top view of the A$_N$DY apparatus 
in the 2011 run is shown in Fig.~\ref{setup}. The HCal modules were positioned 
at a distance of 523~cm from the interaction point, as measured by survey, 
and as close as possible to the beam pipe.

Each HCal consisted of a 12-row $\times$ 9-column matrix
of (10 cm)$^2\times117$-cm long lead cells, each with an embedded
$47\times47$ matrix of scintillating fibers~\cite{Ar98}.  For the 2012 run, 
two 5-row $\times$ 2-column arrays were deployed above and below the beams 
to create an annular HCal with a central $20\times20$ cm$^2$ hole for 
the beams.  The A$_N$DY apparatus also had a pair of 16-element scintillator 
annuli mounted symmetrically about IP2 to serve as a beam-beam counter 
(BBC)~\cite{Bi01}, a pair of 7-row $\times$ 7-column lead glass detector 
arrays serving as small electromagnetic calorimeters (ECal) at a fixed 
(variable) position for the 2011 (2012) run, a scintillator preshower array,
and a pair of zero-degree calorimeter (ZDC) modules~\cite{Ad03} that
faced each beam.  A GEANT~\cite{GEANT} model of A$_N$DY was created, and 
uses inputs from PYTHIA 6.222~\cite{PYTHIA6222}, hereafter referred to as full simulation. 

Polarized proton collisions (i.e., $p^{\uparrow}+p^{\uparrow}$) were initiated at IP2 at systematically
different times in stores during the 2011 run to assess the impact on 
operations.  An automated procedure for bringing A$_N$DY into collisions 
was developed, and as it was repeatedly demonstrated, it can be done without 
significant impact on the beam lifetime and luminosities at other interaction 
points. The colliding beam luminosity
at IP2 was measured by the vernier scan technique and resulted in
$\sigma$=0.94$\pm$0.08 mb for the effective cross section of
coincidences between the ZDC modules which were used to continuously monitor the
luminosity. The $A_N$ results reported here were from 6.5 pb$^{-1}$ of integrated 
luminosity during the 2011 run at $\sqrt{s}=500\,$GeV. For the jet cross section, 
we used 2.5 pb$^{-1}$ of integrated luminosity accumulated during the 2012 run 
at $\sqrt{s}=510\,$GeV, since a possibility to move ECal modules away from 
the beam pipe in that run provided an unobstructed view of the HCal. 
The polarization of each beam was
measured by a relative polarimeter at several times in each fill.  The
relative polarimeter was calibrated from measurements from an absolute
polarimeter, resulting in the average polarization $P_{beam}$=0.526$\pm$0.027 
for the Blue beam used in the jet $A_N$ measurements at $x_F>0$ in the run 
2011. The Yellow beam polarization for the jet $A_N$ at $x_F<0$ was 
$P_{beam}$=0.511$\pm$0.028~\cite{polarization}.

The data by A$_N$DY is from 32-channel 70 MHz flash analog-to-digital
(ADC) converters with 0.25 pC/count sensitivity and noise levels
$<\,$0.25~pC.  Online pedestal corrections were made. Pedestal-corrected ADC 
counts were then analyzed by field-programmable gate arrays  (FPGA) to derive 
an event trigger.  The majority of the data were from a jet trigger that 
summed the ADC response from each HCal module, excluding the outer two 
perimeters of cells.  This trigger is sensitive to electromagnetic (EM) 
and hadronic fragments of jets.  Events were also acquired from 
a minimum-bias trigger that required minimum charge (approximately half that
from a minimum-ionizing particle) from any element 
of the annular BBC that faced each beam, sometimes with a collision vertex 
requirement, where the vertex is reconstructed by the FPGA from the measured 
time difference between the two BBC annuli. Some events were acquired when 
either ZDC crossed threshold as a way to tune bunch-crossing scalers 
used to monitor luminosity as well as the polarization of colliding beams.

\begin{figure}
  \centering
  \includegraphics[height=2.56in]{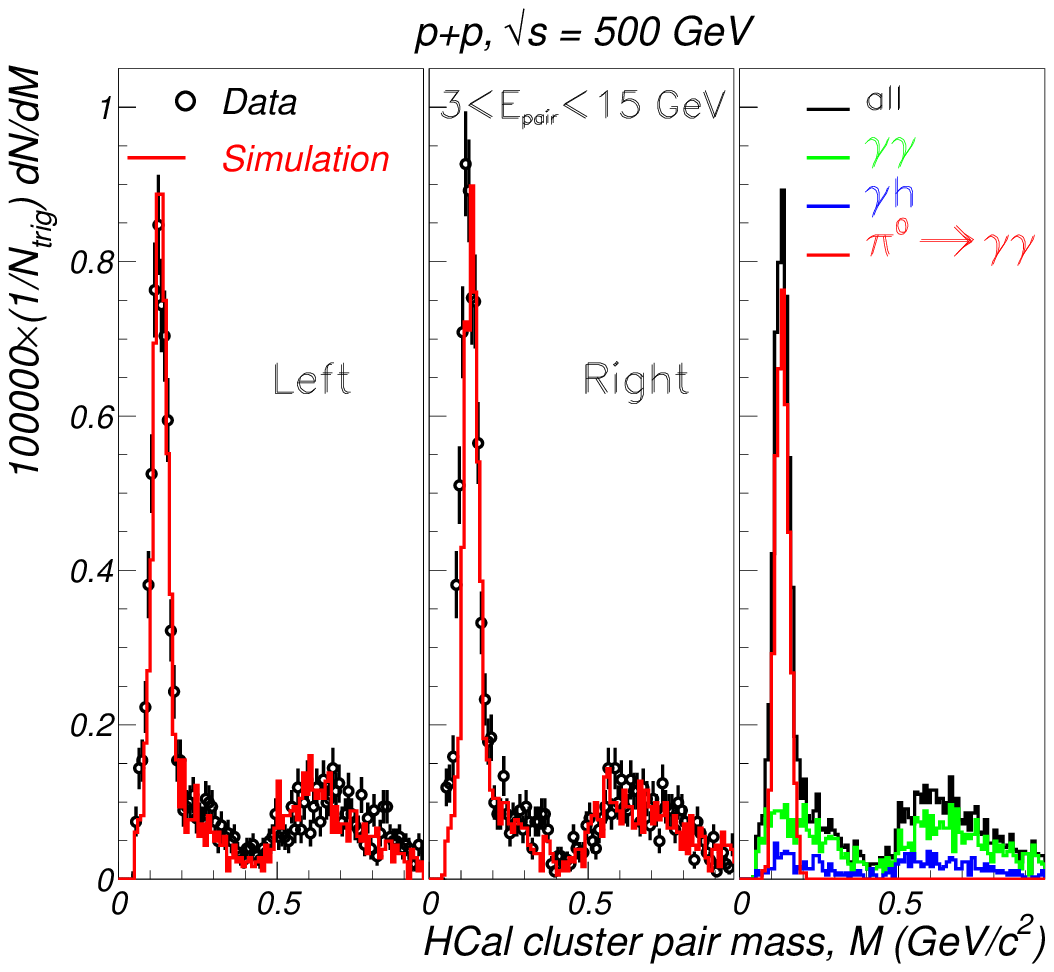}
  \caption{Cluster pair mass distributions, normalized by number of 
  minimum-bias triggers, for single-tower clusters that are primarily photons, 
  showing $\pi^0\rightarrow\gamma\gamma$. 
  (left and middle) Data to simulation comparison for the HCal modules. 
  (right) Association analysis of the simulation showing contributions 
  to the cluster pair mass.}
  \label{mgg}
\end{figure}

The HCal had individual cell gains adjusted prior to colliding beam
operation based on their cosmic-ray muon responses.  The first step in the
offline analysis was to determine the absolute energy scale of the
HCal modules by reconstruction of $\pi^0\rightarrow\gamma\gamma$ from pairs 
of single-cell clusters that had neighboring cells with energy $E'<0.11$~GeV, 
where $E'$ is the incident photon equivalent energy. The reconstructed 
invariant mass of  single-cell cluster pairs, presented in Fig.~\ref{mgg} 
(left, middle) for the two HCal modules, shows an excellent agreement 
between data and simulation. The $\pi^0$ identification was confirmed 
by associating the single-cell clusters reconstructed from our full simulation 
with $\pi^0$ decays generated by PYTHIA, as shown in Fig.~\ref{mgg} 
(right). There is an evident $\pi^0$ peak in the cluster pair mass distribution, 
and backgrounds are mostly photon-photon or photon-hadron combinatorics. 
Offline analysis also refined the relative calibration of all cells by 
a combination of $\pi^0$ reconstruction and the matching of energy deposition 
distributions from single cells between data and full simulation.  For the jet 
analyses described below, the energy calibration was adjusted to account for
the average difference between hadronic and EM showers from full simulation.
We used $E=1.12 \times E'-0.1$~GeV, where $E$ is the equivalent
incident energy measured by individual cells as used in the jet
finders and $E'$ is from $\pi^0$ calibration.  The rescaling of the energy calibration
from neutral pion finding to set the jet-energy scale is further discussed below.

An additional check of the calorimeter hadronic response can be done from reconstruction 
of known mesons or baryons. Evidence for $\rho^0\rightarrow\pi^+\pi^-$ and 
$\Delta\rightarrow N\pi$ was observed in the data, but they were not used 
for calibration because of their large widths. There is also evidence for 
$f_0\rightarrow\pi^+\pi^-$ and $f_2\rightarrow\pi^+\pi^-$, but the mass of 
$f_0$ is not so well known~\cite{PDG} and $f_2$ has a large 
width. Low mass baryons and mesons built from strange quarks 
($\Lambda$, $K_S$) are observed as well. Their utility for calibration 
is impacted by their proximity to the sum of daughter masses and by
their weak decays, resulting in large decay lengths, because nearly all
particles are produced with large Lorentz factor in the forward direction.
The $K^{0*}$ is the natural choice to check hadronic corrections to 
the calibration; since it has small width, it undergoes strong decay
meaning there is no displaced vertex, it is sufficiently more massive
than its daughters, and it is prolifically produced in the forward direction.
The $K^{0*}$ decays with nearly 100\% branching ratio to $K\pi$.

\begin{figure}
  \centering
  \includegraphics[height=2.56in]{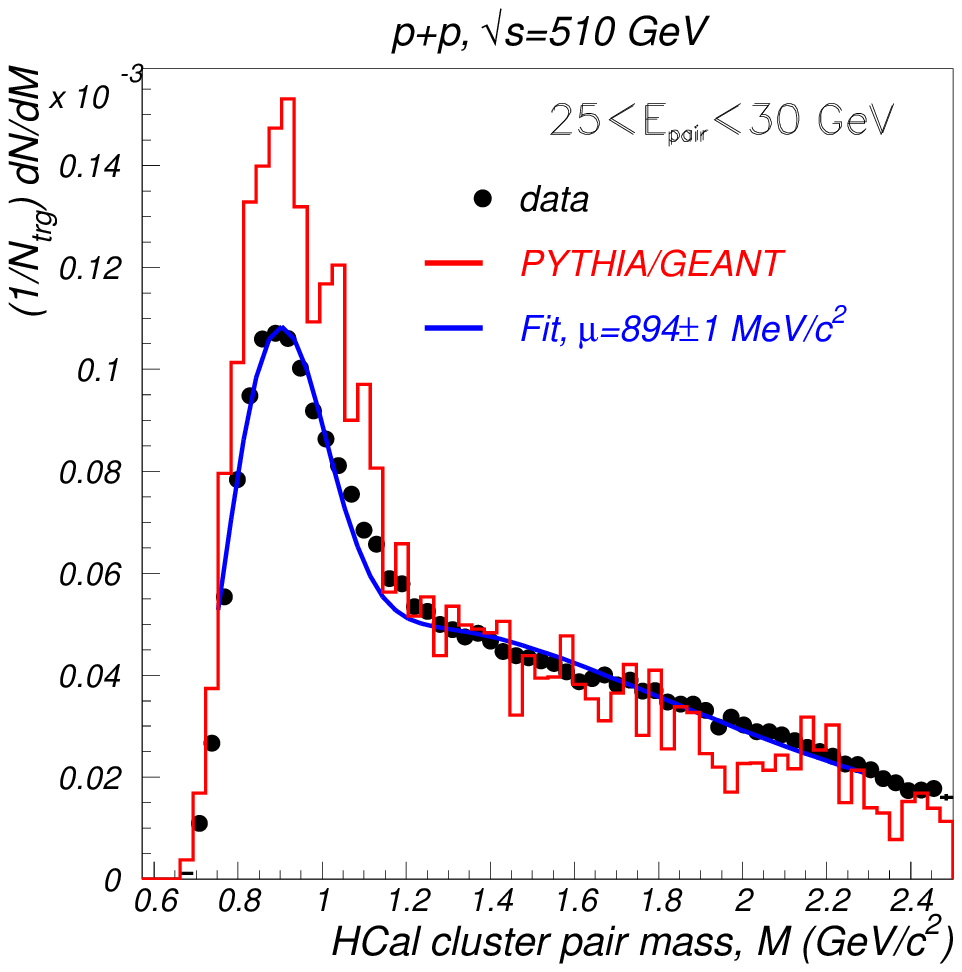}
  \caption{Cluster pair mass distributions, where each cluster is required to have
  energy deposition in the matching BBC detector corresponding to
  a minimum-ionizing particle, from data and simulation.  The peak in the
  data is consistent with the known mass \cite{PDG} of $K^{0*}$, reconstructed
  via $K^{0*}\rightarrow K^-\pi^+$ (and charge conjugates), as determined from
  a fit to the data using a Gaussian peak (centroid, $\mu$) plus background.}
  \label{kstar}
\end{figure}

The reconstruction of $K^{0*}$ is done by clustering the response
of the HCal to an event and choosing ``hadronic-like'' clusters (i.e., clusters
that include multiple towers). The four-momentum of a cluster is then
calculated from the cluster energy, the energy-averaged transverse 
positions ($x, y$) of the cluster and the $z$ position of the collision vertex, 
assuming the cluster is created by a particle originating from the collision 
point, and further assuming the identity of the particle that produced 
the cluster. Photons are a significant background in the HCal when searching 
for particles that decay to charged hadrons. Matching the clusters to energy 
deposition in a BBC detector by assuming a straight-line trajectory of the particle 
from the collision point to the cluster, assists in discriminating charged
particles from photons. Cluster pair mass distributions in Fig.~\ref{kstar} from data and 
from full simulation both show a clear peak attributed to 
$K^{0*}\rightarrow K^-\pi^+$ (and charge conjugates, since charge sign 
is not measured). The pair mass distribution is scaled by the number of
jet triggers, which for full simulation comes from a trigger emulation that
gives a good description of data. The $K^{0*}$ yield is not well modeled by PYTHIA.  
The peak in the data is consistent with the known mass of $K^{0*}$ ($895.8\pm0.2$ MeV/c$^2$ 
\cite{PDG}), as shown by the fit to the data in Fig.~\ref{kstar}.

Our final jet results use the anti-$k_T$ jet algorithm~\cite{Ca08} with a cone
radius of $R_{jet}$=0.7 radians in ($\eta,\phi$) space, although we have also
considered cone algorithms~\cite{No12}.  For each event, the
$\eta_i,\phi_i$ of each cell with $E>E_{thr}$ ($E_{thr}$=0.25 GeV is used to make towers) 
is reconstructed from its surveyed position and the $z$ position of the 
collision vertex for the event. The anti-$k_T$ algorithm reconstructs the jet 
by a pair-wise merging of towers separated by
$d_{ij}=\textrm{min}(k^{-2}_{T,i},k^{-2}_{T,j})\times(R^2_{ij}/R_{jet}^2)$,
when $d_{ij}<1/k^2_{T,i}$ for any $i$.  Each tower has a transverse
momentum $k_{T,i}=E_i/\textrm{cosh}(\eta_i)$, assuming zero mass for
the incident particle.  Pairs of tower clusters are separated by
$R_{ij}=\sqrt{(\eta_i-\eta_j)^2+(\phi_i-\phi_j)^2}$.  The merging
procedure is repeated until all towers are accounted
for.  A valid jet, within a fiducial volume, is considered to have 
$|\eta_{jet}-\eta_0|<\mathrm{d}\eta$ and $|\phi_{jet}-\phi_{0}|<\mathrm{d}\phi$, 
where $\eta_{jet}$ and $\phi_{jet}$ are computed from the energy-weighted
averages of towers included in the jet within the acceptance centered
at $\eta_0,\phi_0$ of half-width d$\eta$, d$\phi$.  For the 2011 data, both
ECal and HCal cells were considered.  For the 2012 data with ECal positioned 
beyond the HCal acceptance, only HCal cells are considered.  Our cross section
and analyzing power results are reported using $\eta_0$=3.25,
d$\eta$=0.25 and d$\phi$=0.5.  The HCal module to the left (right) of
the oncoming beam has $\phi_0=0 \, (\pi)$.

\begin{figure}
  \centering
  \includegraphics[height=2.4in]{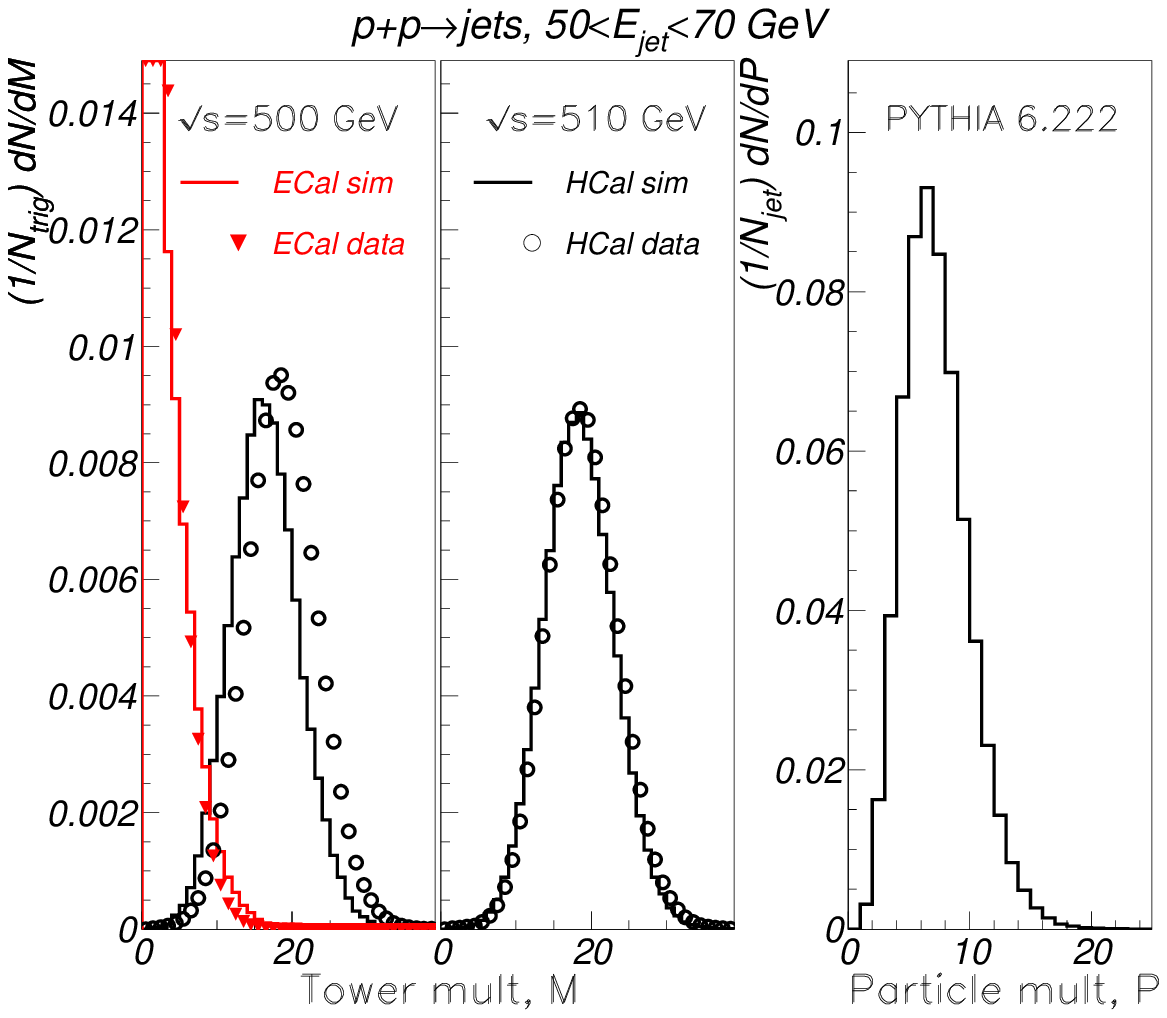}
  \caption{Tower multiplicity distributions for forward jets for data
  compared to full simulation for (left) jets
  from $\sqrt{s}=500\,$GeV collisions,  as used for the jet
  analyzing power; and for (middle) jets from $\sqrt{s}=510\,$GeV collisions, 
  as used for the jet cross section. (right)  Multiplicity of particles 
  produced by PYTHIA 6.222~\cite{PYTHIA6222} that gives rise to the forward jet.}
  \label{mult}
\end{figure}

The tower multiplicity distributions for valid jets are shown in
Fig.~\ref{mult}.  This figure compares jets reconstructed from data to
jets reconstructed from the full simulation of $p+p$ collisions,
where the GEANT response uses the individual cell calibrations to
produce simulated ADC values, and the jet trigger is emulated by the
same algorithm used by the FPGA for our measurements.  In general, the
simulation gives a good description of the data, consistent with
minimal contributions from single-beam backgrounds, as determined from
direct measurement, or from other unknown sources of energy deposition
(underlying event).  Small increases in the HCal multiplicity for 2011 (left panel of Fig.~\ref{mult}) are
attributed to ancillary material (e.g., cables from 
the ECal modules in front of the HCal) not included in GEANT,
prompting us to use the 2012 data for the jet cross section.
The reconstructed jets have a broad tower multiplicity distribution 
whose mean value increases as $E_{jet}$ increases.  Given that data and 
full simulation agree, we can then infer the distribution of particles in 
the jet by applying the anti-$k_T$ jet finder to detectable particles 
produced by PYTHIA. These particle jets have similar multiplicity to those 
reconstructed in fixed-target hadroproduction experiments~\cite{ftjet}.  Such low 
multiplicity jets are generally not accessible in hadron colliders because 
their $p_T$ is too low. Forward detection at large magnitude $x_F$ makes 
these measurements possible.

\begin{figure}
  \begin{tabular}{ll}
  \includegraphics[height=2.4in]{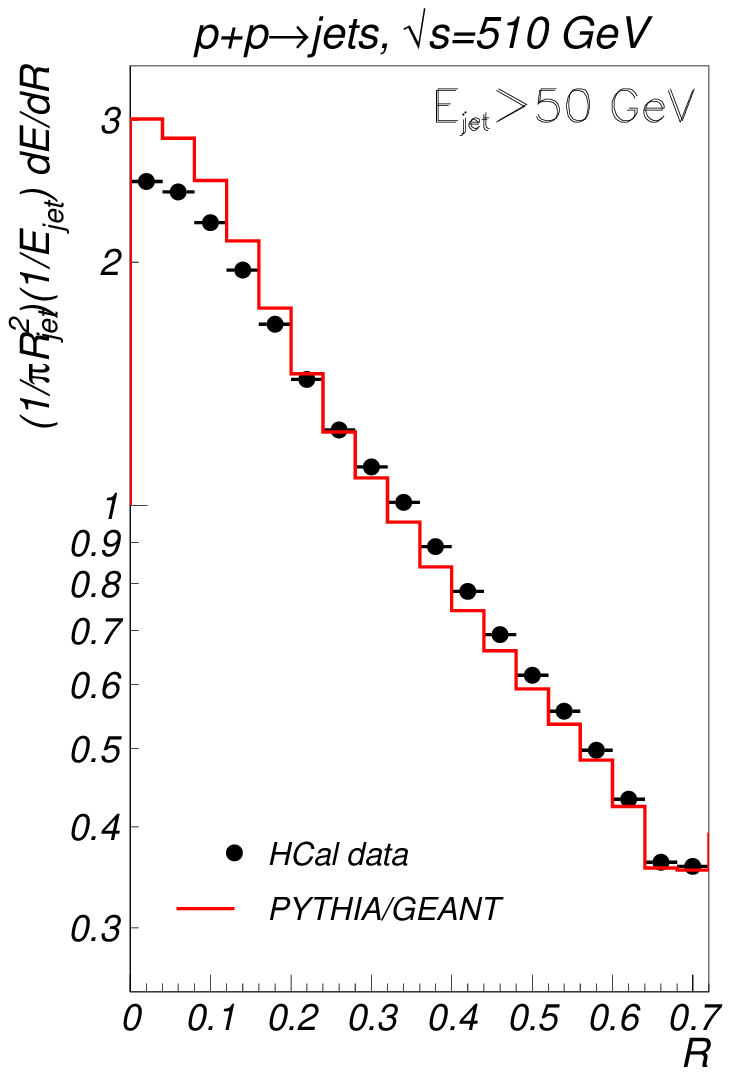} &
  \includegraphics[height=2.4in]{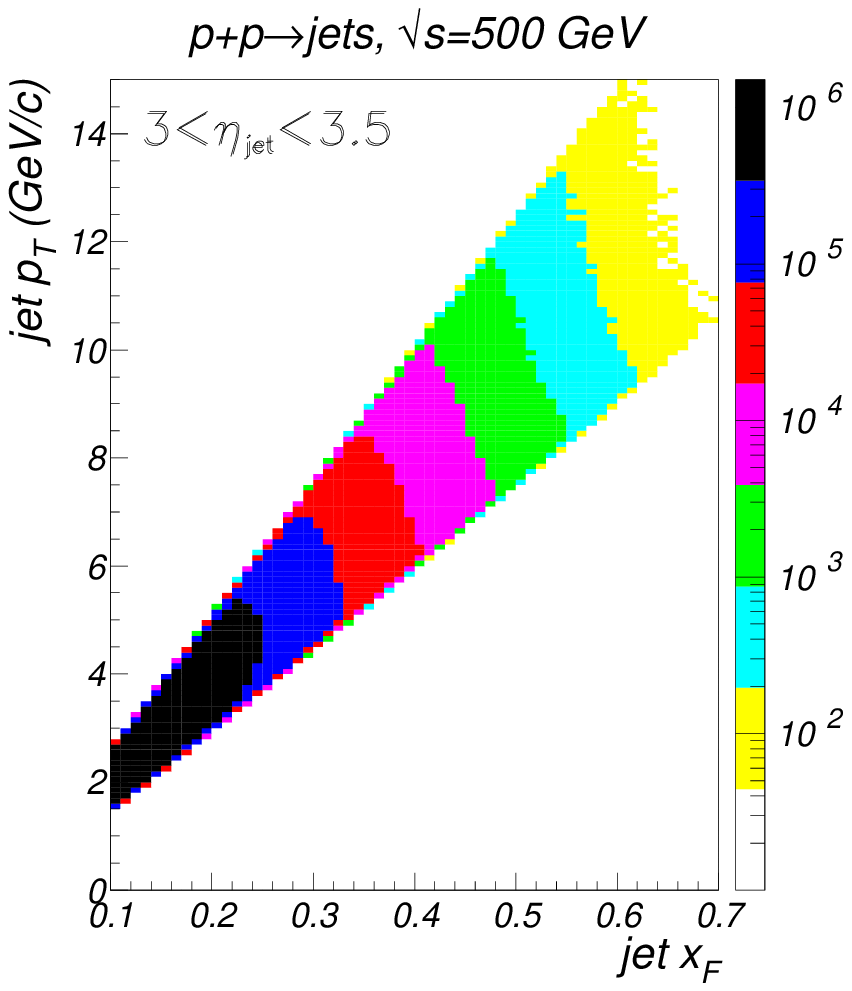}
  \end{tabular}
  \caption{(left) Event averaged jet shape, corresponding to how the energy
  depends on R, the distance of a tower from the thrust axis in
  ($\eta,\phi$) space. (right) Correlation between jet $x_F$ and $p_T$.
  The color scale is the number of events.}
  \label{shape}
\end{figure}

The towers included in the jets have their energy
distributed relative to the thrust axis in a manner that is typical of
a jet (Fig.~\ref{shape}).  Most of the energy is concentrated near the
thrust axis.  As towers become increasingly distant from the thrust
axis, on average they contribute little to the energy of the jet.  The
data is well described by our full simulation, although there
are some indications that jets produced by PYTHIA 6.222~\cite{PYTHIA6222} have energy
concentrated closer to the thrust axis relative to our measurements.
Also shown in Fig.~\ref{shape} is the correlation between $x_F$
and $p_T$ for the jet events.  The d$\eta$ requirement strongly correlates
these two kinematic variables.

\begin{figure}
  \centering
  \includegraphics[height=2.5in]{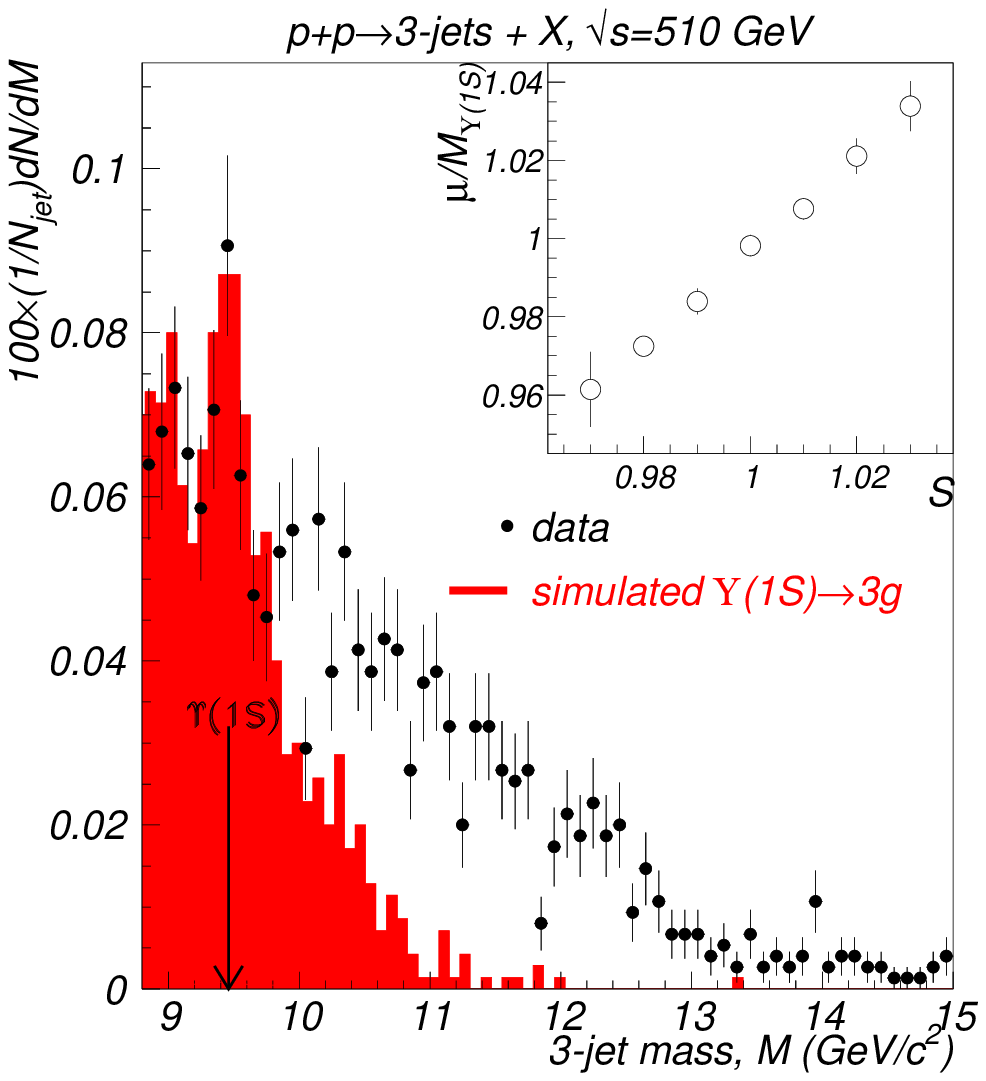}
  \caption{Test of jet energy scale from 3-jet mass.  For the inset, $S$
  rescales the jet energy, $\mu$ is the peak centroid, and
  $M_{\Upsilon{\rm (1S)}}$ is the known mass \cite{PDG}. Simulation is
  the single channel $\Upsilon$(1S)$\rightarrow 3g$ using GEANT, for
  particles produced by the PYONIA generator~\cite{PYTHIA6425}.}
  \label{3jetmass}
\end{figure}

The jet energy scale was established by comparing tower jets reconstructed
from the full simulation to particle jets reconstructed from
PYTHIA, and resulted in the hadronic compensation described earlier.
A check of the energy scale was made for 3-jet events in the data by
observation of a narrow structure in 3-jet mass distribution attributed 
to the 3-gluon decay of $\Upsilon$(1S)~\cite{CLEO}. 
Fig.~\ref{3jetmass} shows the 3-jet mass distribution compared to 
the simulated $\Upsilon$(1S)$\rightarrow 3g$. The peak has statistical 
significance of $3.5\sigma$. The centroid of the peak depends smoothly 
on $R_{jet}$ used in the anti-$k_T$ algorithm, as do our measures of the jet 
energy scale from simulation. The mass peak is narrow because it comes from 
the jets that consist primarily of photons, electrons, and positrons, 
as deduced from the simulation. The uncertainty of the jet energy scale is 
constrained by the variation of the mass peak centroid ($\mu$) scaled 
by the known mass ($M_{\Upsilon{\rm (1S)}}$ \cite{PDG}) with $S$, as shown 
in the inset to Fig.~\ref{3jetmass}.  The value of $S$ rescales the energies of towers 
considered by the jet finder.  The jet-energy scale variations probe the 
modification of the HCal calibration deduced from neutral pion reconstructions.

The forward jet production cross section was measured by scaling the
number of reconstructed jets by the measured integrated luminosity and
correction factors described here.  For jet triggers, there is a
trigger efficiency ($\epsilon_{trig}$) dominated by the variation of
$\eta$ along the collision vertex distribution.  The efficiency $\epsilon_{trig}$ is
evaluated as a function of jet energy from the full simulation, and is
checked by comparing invariant jet cross sections from jet-triggered
events to cross sections determined from the minimum-bias trigger.
The jet detection efficiency ($\epsilon_{jet}$) is determined from the
ratio of number of tower jets within the acceptance to the number of
particle jets within the acceptance, and resulted in the value of 0.83
independent of the jet energy.  The value of $\epsilon_{jet}$ 
is checked by systematically varying the acceptance and assessing
the stability of the resulting invariant cross section.  Sources of
systematic uncertainty are (a) values of $\epsilon_{trig}$ and
$\epsilon_{jet}$; (b) time-dependent effects from either HCal gain
stability or from beam conditions; (c) jet-energy scale uncertainties;
(d) luminosity normalization; and (e) jet-finder parameters
($R_{jet},E_{thr}$) that also probe underlying event contributions.

\begin{figure}
  \centering
  \includegraphics[height=2.78in]{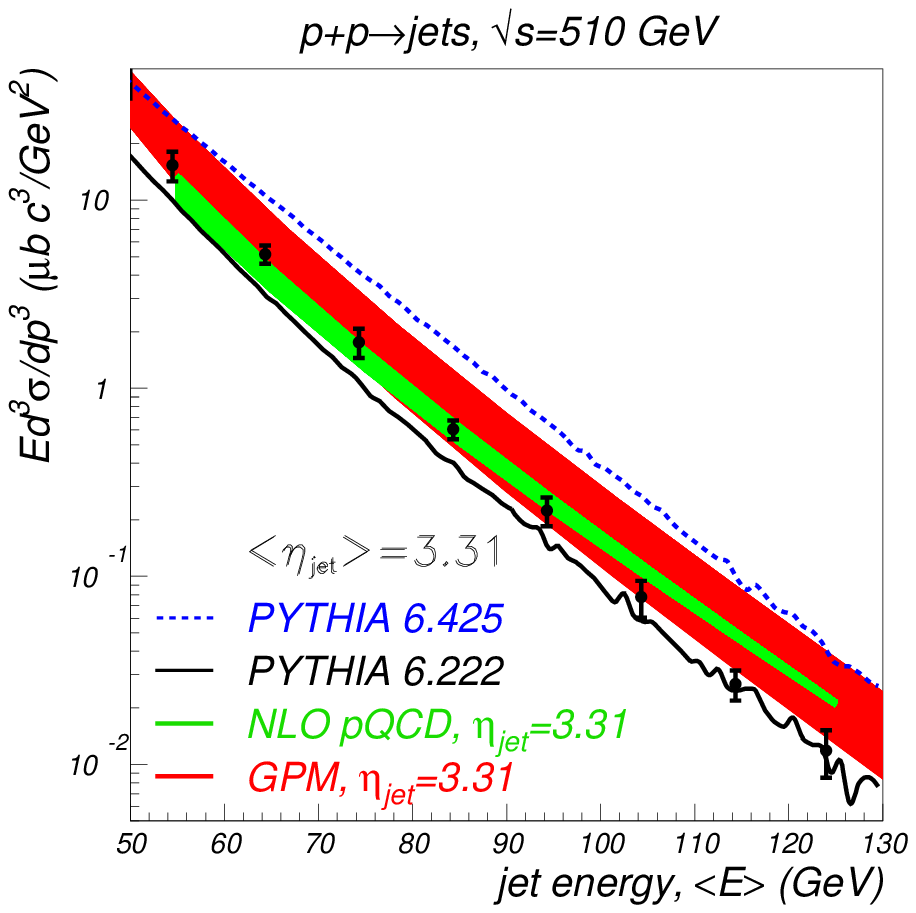}
  \caption{Invariant forward jet cross section compared to predictions
  by PYTHIA, next-to-leading order pQCD calculations, and the
  generalized parton model.  The error bars include systematic
  uncertainties, described in the text.}
  \label{jetcs}
\end{figure}

The correlation between tower-jet energy and particle-jet energy from
full simulation also addresses jet-energy resolution.  The distributions
of tower-jet energy in bins of particle-jet energy are found to be described
by Gaussian functions.  Jet-energy resolution is deduced from the ratio
of the fitted sigma and centroid.  This ratio yields 
$\delta E_{jet}/E_{jet}\approx$ 16\%, independent of jet energy.  The impact of
jet-energy resolution is accounted for in our jet cross sections through
$\epsilon_{jet}$.

\begin{figure}
  \centering
  \includegraphics[height=2.78in]{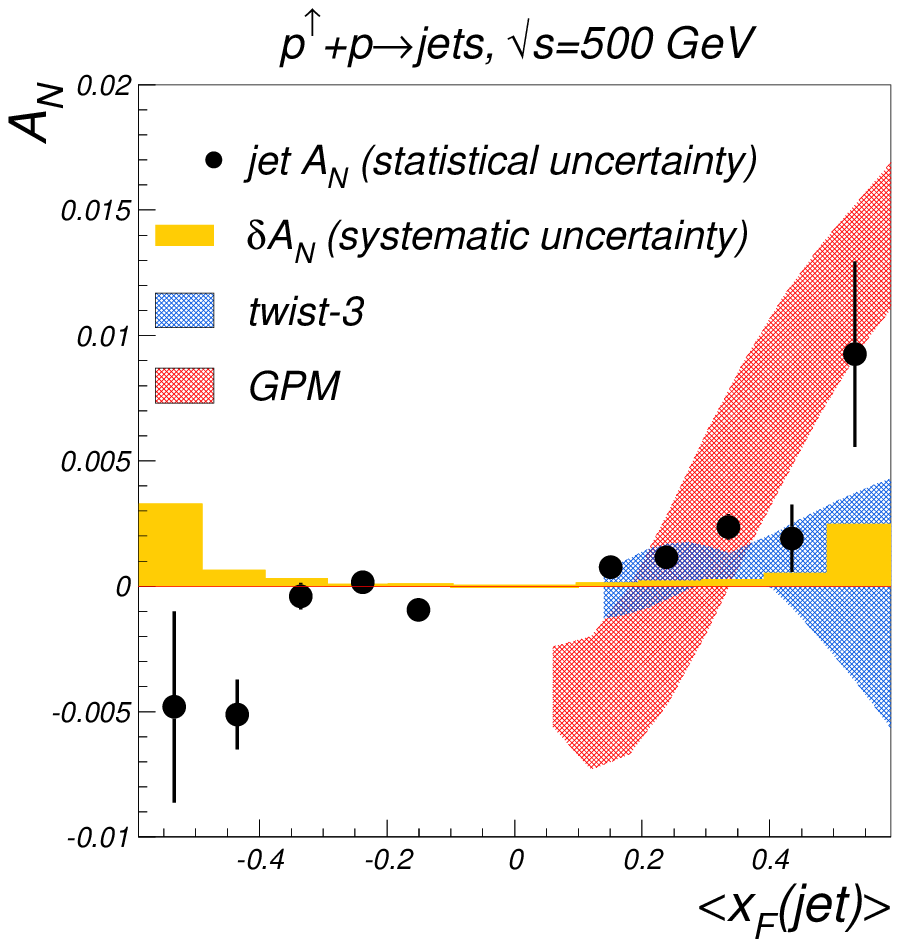}
  \caption{Analyzing power for forward jet production compared to theoretical 
  model calculations. Jets are reconstructed with the anti-$k_T$ algorithm 
  using $R_{jet}$=0.7. Preliminary results \cite{No12} reported comparable 
  $A_N$ with the mid-point cone algorithm.  Systematic uncertainty estimates
  are described in the text, and do not include scale uncertainty from
  the beam polarization measurements.  Theoretical systematic uncertainties
  are described in Refs \cite{Ga13} and \cite{Ans13}.}
  \label{anjet}
\end{figure}

The resulting distribution of forward jet cross section as a function
of energy is compared to next-to-leading order (NLO) pQCD
calculations~\cite{Vogel12} and calculations from the generalized
parton model (GPM) \cite{Ans13} in Fig.~\ref{jetcs}. The
cross section is averaged over the acceptance, resulting in 
$\langle\eta_{jet}\rangle$=3.31. Both theoretical calculations provide
a fair description of the cross section, supporting the conclusion
that forward jets originate from hard scattering. The error bands for
the calculations reflect the scale dependence, with the upper limit of
each band using scale $\mu=p_T/2$ and the lower limit using
$\mu=2p_T$.  The scale dependence of the GPM calculations is larger
than for NLO pQCD, likely reflecting that the GPM is a leading-order
calculation, albeit with parton distribution functions that depend on
transverse momentum that are constrained to fit unpolarized SIDIS data
with factorization assumed.  Our results are also compared to PYTHIA
6.222~\cite{PYTHIA6222} and 6.425~\cite{PYTHIA6425} predictions for
anti-$k_T$ jets reconstructed from stable particles that are within
the detector acceptance.  In Ref.~\cite{No12} it was shown that PYTHIA 6.222
predicts that forward jets arise from partonic hard scattering.  Analysis
of particle jets generated by PYTHIA 6.222 show that the forward jet $x_F$
is strongly correlated with the Bjorken-$x$ of the parton (most likely
a valence quark) from the proton with $p_z>0$.  The Bjorken-$x$ distributions
in bins of $x_F$ can be described by Gaussian functions with $\sigma$=0.06 at
$x_F$=0.2 increasing to $\sigma$=0.09 at $x_F$=0.4.  There is no
correlation between jet $x_F$ and the Bjorken-$x$ of the parton from
the other proton, resulting in a broad distribution of parton $x$
values that extends down to $10^{-4}$, as is expected for forward
particle production.  The low-$x$ part of that distribution can be
accessed by detecting forward dijets.  PYTHIA 6.222 precedes tunings
based on Tevatron data which resulted in later versions (e.g., 6.425)
used by the LHC.  Versions of PYTHIA that predate tunings for the LHC
are known to accurately describe large $x_F$ $\pi^0$
production~\cite{Bl04}, and are known to lose accuracy for more
complicated multi-particle correlations~\cite{Go09}.

The forward jet $A_N$ is measured by the cross-ratio method from yields in the 
nominally mirror symmetric beam-left and beam-right HCal modules, sorted by 
the polarization direction of the Blue beam heading towards the detector 
and averaged over the polarization direction of the opposite beam (Yellow) 
for positive $x_F$ (and vice versa for negative $x_F$):
\begin{equation}
A_N=\frac{1} {P_{beam}}
\frac {\sqrt{N_L^{\uparrow}N_R^{\downarrow}}-\sqrt{N_L^{\downarrow}N_R^{\uparrow}}}
      {\sqrt{N_L^{\uparrow}N_R^{\downarrow}}+\sqrt{N_L^{\downarrow}N_R^{\uparrow}}},
\end{equation}
where $N_{L(R)}^{\uparrow(\downarrow)}$ is the number of jet 
events in the beam-left (-right) module for the spin direction up (down). 
This method cancels systematics, such as luminosity and detector asymmetries, 
through second order. Each fill has a pattern of spin directions for bunches 
of beam injected into RHIC.  A specific crossing of bunches from the two rings
is the remainder after dividing the RHIC clock count for an event by 120.
The bunch-crossing distribution has characteristic holes that correspond 
to missing bunches from one or the other beam.  The pattern of polarization
directions for that fill recorded at A$_N$DY originating from information
broadcast by RHIC is then used to accumulate $N_{L(R)}^{\uparrow(\downarrow)}$ 
in the analysis.  Since the RHIC broadcast information specifies polarization 
directions at the polarized ion source, we rely on the measurement of spin 
asymmetries for far-forward neutron production measured by the ZDC, where 
the $A_N$ was previously measured~\cite{Fu07}, to ensure the jet $A_N$ 
is measured with the proper sign.

\begin{table}
\centering
\caption{$A_N$ for forward jet production at $\sqrt{s}=500$~GeV.}
\vspace{0.1cm}
\begin{tabular}{rcrcrcr}
\toprule
~~$\langle x_F\rangle$ &~~~~& $A_N$~~~&~~~~& $\delta_{A_N}^{stat}$~~ &~~~~& $\delta_{A_N}^{syst}~~$\\ 
\midrule
 $-0.534$ & & $-0.00481$ & & 0.00381 & & 0.00324 \\
 $-0.434$ & & $-0.00511$ & & 0.00139 & & 0.00059 \\
 $-0.335$ & & $-0.00039$ & & 0.00053 & & 0.00027 \\
 $-0.238$ & &   0.00017  & & 0.00023 & & 0.00004 \\
 $-0.151$ & & $-0.00094$ & & 0.00016 & & 0.00006 \\
   0.151  & &   0.00077  & & 0.00016 & & 0.00009 \\
   0.238  & &   0.00116  & & 0.00022 & & 0.00016 \\
   0.335  & &   0.00237  & & 0.00052 & & 0.00022 \\
   0.434  & &   0.00192  & & 0.00135 & & 0.00048 \\
   0.534  & &   0.00926  & & 0.00370 & & 0.00242 \\
\bottomrule
\end{tabular}
\label{antab}
\end{table}

Our measured forward jet $A_N$ is shown in Fig.~\ref{anjet} compared to 
twist-3 pQCD calculations~\cite{Ga13} and GPM
calculations~\cite{Ans13}, and presented in Table~\ref{antab}. 
Non-zero $A_N$ for forward jets is expected for the Sivers effect, but 
not for spin-dependent fragmentation effects because the jet finding integrates over 
the produced hadrons.  The measured jet $A_N$ at $x_F>0$ is small and positive. 
There is a hint of a negative $A_N$ at $x_F<0$.  
One check for systematic effects was to fit the spin asymmetry
($\epsilon=P_{beam}A_N$) measured in each jet $\langle x_F\rangle$ bin
for each RHIC fill by a constant.  The resulting $\chi^2$ per degree
of freedom from these fits is close to unity, and is consistent with
the statistical uncertainties, meaning the systematic uncertainties are small.  A
more quantitative check for systematic effects was to establish if an
effectively unpolarized sample of $p+p$ collisions had $A_N$
consistent with zero.  This was accomplished by a random reversal of
the spin direction for half of the filled bunch crossings.  The mean
value of $\epsilon$ for $\sim\,$100 random spin direction patterns had
values $10^{-5}<\epsilon<10^{-4}$ resulting in the systematic uncertainty 
estimate of $2\times10^{-4}$ for the jet $A_N$.  The systematic uncertainties 
of $A_N$ are estimated by varying the jet finder and valid jet parameters.  
Our jet $A_N$ measurement is limited by statistics. The measured small 
and positive jet $A_N$ is naively expected because 
$A_N(\pi^+){\approx}-A_N(\pi^-)$, thus giving cancelling contributions 
from $\pi^\pm$ in a jet. 

Comparisons of our measured forward jet $A_N$ to theory have already
been discussed in Refs. \cite{Ga13, Ans13}.  A few key aspects of this
comparison are presented here.  Both the GPM and the twist-3 pQCD
calculations fit the Sivers function to transverse single-spin
asymmetries from SIDIS.  It is important to recognize that the
Bjorken-$x$ range of the SIDIS data has little kinematic overlap with
either forward jet data or forward pion data \cite{STARpi0}.  Unlike
the case for pion production, both the GPM and the twist-3 pQCD
calculation agree that the forward jet $A_N$ should be small and
positive.  Their phenomenological extractions of the Sivers function
from SIDIS are compatible with the sign and magnitude of $A_N$ in
$p^{\uparrow}p\rightarrow $ jet + $X$.  Neither calculation considers
negative $x_F$ jet production.  Other theoretical work
\cite{Zh14,Be14} involving tri-gluon correlators and low-$x$ phenomena
address transverse single spin effects at negative $x_F$.  A future
measurement that improves the precision of these measurements is
required to compare to theory for possible spin effects at negative $x_F$.

In conclusion, we have made first measurements of forward jet
production in $p{^\uparrow}+p$ collisions at $\sqrt{s}=500\,$GeV.  
Our measured cross section is consistent with dominant
contributions from partonic hard scattering, even though the transverse
momentum for the produced jets is small ($2<p_T<10$ GeV/$c$).  We have
measured the analyzing power for forward jet production, and find it
to be small and positive.  Our measurements constrain knowledge~\cite{Ga13} 
of Sivers functions, that are related to parton OAM through models.  
It remains the case that the most definitive experiment to test present 
understanding is a measurement of the analyzing power for Drell-Yan production.
        
We thank the RHIC Operations Group at BNL.  This work was supported in part
by the Office of NP within the U.S. DOE Office of Science (contract DE-SC0012704), the Ministry of Ed.
and Sci. of the Russian Federation, and the Ministry of Sci., Ed. and Sports
of the Rep. of Croatia, and IKERBASQUE and the UPV/EHU under program UFI 11/55.

%%\end{linenumbers}


\begin{thebibliography}{99}

\bibitem{kfliu}
           K-F.~Liu, \Journal{\NPA}{928}{99}{2014}.

\bibitem{E704}
           B.~E.~Bonner {\it et al.}, \Journal{\PRL}{61}{1918}{1988};
           D.~L.~Adams {\it et al.}, \Journal{\PLB}{261}{201}{1991}; 
           {\bf 264}, 462 (1991).

\bibitem{Ka78}
           G.~L.~Kane, J.~Pumplin, W.~Repko,
           \Journal{\PRL}{41}{1689}{1978}.

\bibitem{Si90}
           D.~Sivers, \Journal{\PRD}{41}{83}{1990}; {\bf 43}, 261
           (1991).

\bibitem{Co93}
           J.~Collins, \Journal{\NPB}{396}{161}{1993}.


\bibitem{Bu06}
           M.~Burkardt, G.~Schnell, \Journal{\PRD}{74}{013002}{2006};
           A.~Bacchetta, M.~Radici, \Journal{\PRL}{107}{212001}{2011}. 

\bibitem{lowE}
           R.~D.~Klem {\it et al.}, \Journal{\PRL}{36}{929}{1976};
           W.~H.~Dragoset {\it et al.}, \Journal{\PRD}{18}{3939}{1978};
           B.~E.~Bonner {\it et al.}, \Journal{\PRD}{41}{13}{1990};
           C.~E.~Allgower {\it et al.}, \Journal{\PRD}{65}{092008}{2002}.

\bibitem{STARpi0}
           J.~Adams {\it et al.},
           \Journal{\PRL}{92}{171801}{2004};
           B.I.~Abelev {\it et al.},
           \Journal{\PRL}{101}{222001}{2008}.

\bibitem{BRAHMS}
           I.~Arsene {\it et al.}, \Journal{\PRL}{101}{042001}{2008}.

\bibitem{STARcs}
           J.~Adams {\it et al.},
           \Journal{\PRL}{97}{152302}{2006}.

\bibitem{SIDIS} 
           A.~Airapetian {\it et al.} (HERMES),
           \Journal{\PRL}{103}{152002}{2009}; {\bf 94}, 012002 (2005);
           C.~Adolph {\it et al.}  (COMPASS), \Journal{\PLB}{717}{376}{2012}; {\bf 717},  383 (2012).

\bibitem{ETQS}
	 A.~V.~Efremov and O.~V.~Teryaev, Sov.~J.~Nucl.~Phys. {\bf 36}, 140 (1982) [Yad.~Fiz. {\bf 36}, 242 (1982)]; 
	 \Journal{\PLB}{150}{383}{1985}; 
	 J.-Qiu and G.~F.~Sterman, \Journal{\PRL}{67}{2264}{1991}; 
	 \Journal{\NPB}{378}{52}{1992}; \Journal{\PRD}{59}{014004}{1998}.

\bibitem{twist3SSA}
	  Y.~Koike, \Journal{\NPA}{721}{364}{2003};
	  C.~Kouvaris {\it et al.}, \Journal{\PRD}{74}{114013}{2006}. 
	 
\bibitem{sign-mismatch}
	  Z.-B.~Kang  {\it et al.}, \Journal{\PRD}{83}{094001}{2011}.

\bibitem{KKMP}
	  K.~Kanazawa {\it et al.}, \Journal{\PRD}{89}{111501}{2014}.

\bibitem{Ar98}
           T.~A.~Armstrong {\it et al.},
             \Journal{\NIMA} {406}{227}{1998}.

\bibitem{Bi01}
           R.~Bindel {\it et al.}, \Journal{\NIMA}{474}{38}{2001}.

\bibitem{Ad03}
           C.~Adler {\it et al.}, \Journal{\NIMA} {499}{433}{2003}.

\bibitem{GEANT} 
           GEANT 3.21, CERN program library.

\bibitem{PYTHIA6222}
           T.~Sj\"ostrand {\it et al.},
           Computer Physics Commun. {\bf 135}, 238 (2001).

\bibitem{polarization}
           RHIC Polarimetry Group, RHIC/CAD Accelerator Physics Note {\bf 490} (2013).

\bibitem{Ca08}
           M.~Cacciari, G.~P.~Salam, G.~Soyez,
           \Journal{\JHEP}{0804}{063}{2008}.

\bibitem{No12}
           L.~Nogach (for A$_N$DY), {\it 20$^{th}$ International
           Symposium on Spin Physics} (2012) [arXiv:1212.3437].

\bibitem{CLEO}
           M.~S.~Alam {\it et al.} (CLEO), \Journal{\PRD}{56}{17}{1997}.

\bibitem{ftjet}
           C.~Bromberg {\it et al.}, \Journal{\PRL}{38}{1447}{1977}; 
           M.~D.~Corcoran {\it et al.}, \Journal{\PRL}{41}{9}{1978}; {\bf 44},  514 (1980);
           M.~W.~Arenton {\it et al.}, \Journal{\PRD}{31}{984}{1985}.

\bibitem{PDG}
           J.~Berringer {\it et al.} (Particle Data Group), \Journal{\PRD}{86}{010001}{2012}. 

   
\bibitem{Vogel12}
           A.~Mukherjee and W.~Vogelsang, \Journal{\PRD}{86}{094009}{2012}. 

\bibitem{PYTHIA6425}
           T.~Sj\"ostrand, S.~Mrenna, P.~Skands,
           JHEP {\bf 0605}, 026 (2006).

\bibitem{Bl04}
           L.~C.~Bland (for STAR), {\it X$^{th}$ Advanced Research
           Workshop on High Energy Spin Physics}, (2003)
           [hep-ex/0403012].

\bibitem{Go09} A.~Gordon (for STAR), {\it Recontres de Moriond QCD and
           High Energy Interactions} (2009) [arXiv:0906.2332].

\bibitem{Fu07} Y.~Fukao {\it et al.}, \Journal{\PLB}{650}{325}{2007}.

\bibitem{Ga13} L.~Gamberg, Z.~-B.~Kang, A.~Prokudin, \Journal{\PRL}{110}{232301}{2013}.

\bibitem{Ans13}
           M.~Anselmino {\it et al.}, \Journal{\PRD}{88}{054023}{2013}.

\bibitem{Zh14} J. Zhou, Phys. Rev. D {\bf 89}, 074050 (2014).

\bibitem{Be14} H. Beppu, K. Kanazawa, Y. Koike, S. Yoshida,
  Phys. Rev. D {\bf 89}, 034029 (2014).

\end{thebibliography}
\end{document}